\documentclass[manuscript]{rmaa}
\usepackage{paralist}
\usepackage[latin1]{inputenc}

\title{Testing the isotropy of the Hubble expansion} 

\author{
  K. Migkas,\altaffilmark{1} 
  M. Plionis\altaffilmark{2,3,4}}

\altaffiltext{1}{Argelander Institute for Astronomy (AIfA), University
  of Bonn, Auf dem Hugel 71, 53121 Bonn, Germany}
\altaffiltext{2}{Astrophysics, Astronomy \& Mechanics Sector, Physics
  Department, Aristotle University of Thessaloniki, 54124 Greece}
\altaffiltext{3}{IAASARS, National Observatory of Athens, 11810 Greece}
\altaffiltext{4}{Instituto Nacional de Astrof\'isica, \'Optica y
  Electr\'onica, Puebla, M\'exico C.P. 72840}

\shortauthor{Migkas \& Plionis}
\shorttitle{Isotropy of Hubble expansion?}

\listofauthors{K. Migkas \& M. Plionis}
\indexauthor{Migkas, K.}
\indexauthor{Plionis, M.}

\abstract{
We have used the {\em Union2.1} SNIa compilation to search for possible 
Hubble expansion anisotropies, dividing the sky in 9 solid angles
containing roughly the same number of SNIa, as well as in the two Galactic hemispheres.
We identified only one sky region, containing 82 SNIa
($\sim$15\% of total sample with $z>0.02$), that
indeed appears to share a significantly different Hubble expansion than the
rest of the sample. 
However, this behaviour appears to be attributed to the joint
``erratic'' behaviour of only three SNIa 
and not to an anisotropic expansion.
We also find that the northern and southern galactic hemispheres have
different cosmological
parameter solutions but still not significant enough
to assert the detection of a Hubble expansion anisotropy.
We conclude that 
even a few outliers can have such an effect as to induce
artificial indications of anisotropies, when the number
of analysed SNIa is relatively small.}

\resumen{

Hemos utilizado la compilación {\em Union2.1} de SNIa 
para buscar posibles anisotropías de la expansión de Hubble, 
dividiendo el cielo en 9 ángulos sólidos que contienen 
más o menos el mismo número de SNIa, así como
en los dos hemisferios galácticos.
Como resultado se identificó una región del cielo que contiene 82 SNIA
(15\% del total con $z>0.02$) y que
parece tener un comportamiento de expansión Hubble
significativamente diferente de resto de la muestra. 
Pero la cause es un efecto sistemático relacionado mayormente al
comportamiento ``errático'' de solo tres SNIa.
Además, el análisis por separado del los dos hemisferios galácticos resulta 
en diferentes parámetros cosmologicos,
pero que todavía no es lo suficientemente significativo
para afirmar la detección de una anisotropía de la expansión Hubble.
Llegamos a la conclusión de que 
incluyendo en el análisis unos pocos SNIa con valores atípicos
puede proporcionar indicaciones artificiales de anisotropías.}

\addkeyword{(cosmology:) Cosmological parameters}
\addkeyword{supernovae: general}
\addkeyword{methods: statistical}

\begin{document}

\maketitle

\section{Introduction}
The isotropy and homogeneity of the Universe, the principle on
which the Friedmann-Robertson-Walker cosmological models are
based, is strongly
supported by several observational data, among which the isotropy of
the Cosmic Microwave Background (CMB) radiation (eg., Ade et
al. 2014), the large scale distribution of radio sources
(Wu, Lahav \& Rees 1999) and the Hubble expansion as traced by 
supernovae type Ia (SNIa)  (eg., Riess et al. 1998, Perlmutter et
al. 1999, Suzuki et al. 2012, Betoule et al. 2014). 

The SNIa are excellent cosmological probes of the
Hubble expansion up to high redshifts ($z\leq 1.7$) due to the fact
that they are considered as "standard candles" (Kowal 1968, Barbon
et al. 1973, Riess et al. 1998, Perlmutter et al. 1999), once the
so-called ``stretch'' and color 
corrections have been applied. These corrections are necessary because
the SNIa peak
brightness correlates with their color, the light-curve
width and the mass of the host galaxy (see Kowalski et al. 2008, Amanullah
et al. 2010, Suzuki et al. 2012, Betoule et al. 2014).  Other
high-$z$ tracers of the Hubble expansion have been proposed, such as
HII galaxies (Melnick, Terlevich \& Terlevich 2000; Plionis et
al. 2011) and GRB's (Ghirlanda et al. 2006). Such cosmic tracers can
be used to test the Cosmological Principle by confirming or not the
isotropy of the Hubble expansion. 
Several recent studies have focused in this subject, among which those of
Kolatt \& Lahav (2001),  Bonvin, Durrer \& Kunz (2006), Schwarz \&
Weinhorst (2007), Blomqvist, Moertsell \& Nobili (2008), Gupta,
\& Saini (2010), Cooke \& Lynden-Bell (2010), Antoniou \&
Perivolaropoulos (2010), Mariano \& Perivolaropoulos (2012),
Kalus et al. (2013), 
Yang, Wang \& Chu (2013), Heneka, Marra \& Amendola (2014), 
Javanmardi et al. (2015).
It is interesting to note that various of the previously
mentioned studies have found a low-significance dipole correlated with
the general direction of the CMB dipole.
 
In this work we use the \textit{Union2.1} sample in order
to search for possible anisotropies of the Hubble expansion. The
approach that we follow is to attempt to identify solid angles that
show a different expansion behaviour with respect to the
rest.
In our analysis we use only those SNIa with redshifts 
$z\geq 0.02$ in order to avoid uncertainties in their estimated distances due
to the local bulk flows (eg., Ma \& Pan 2014; Appleby, Shafieloo \& Johnson
2015).

\section{Data and methodology}

\subsection{Union2.1 SNIa sample}

The \textit{Union2.1} is one of the largest compilations of
supernovae of type Ia (Suzuki et al. 2012). 
It originally consisted of 833 SNe drawn from 19 different datasets,
but after a variety of homogenization selection criteria
(eg. lightcurve quality cuts) 580 remained, out of which 
546 SNIa with $z\geq0.02$ and 29 with $z>1$.
The high-redshift SNIa are extremely important in calculating the
cosmological parameters, because the differences
of various Dark Energy (DE hereafter) models are larger and more
significant at such redshifts (for example see Fig.1 in Plionis et
al. 2011).

We note that the final {\em Union2.1} SNIa distance moduli provided
and used in the current paper have been obtained by assuming a common,
independent of their direction, correction for the ``stretch'', color
and host-galaxy parameters for all SNIa (Suzuki et al. 2012).

In order to visualize the angular and redshift distribution of the
{\em Union2.1} SNIa sample, we produce a pie diagram with the redshift, $z$, 
being the radius and the right ascension, $\alpha$, being the angle
(Fig. 1).
As we can see the higher-$z$ SNIa  are detected preferentially along specific
directions, some of which contain more data than others. This non
uniform distribution of data, in terms of coordinates, is a result of
the fact that the observational campaigns cover very small solid
angles of the sky. 
\begin{figure}[!t]
\center
 \includegraphics[width=0.99\textwidth,height=9.5cm]{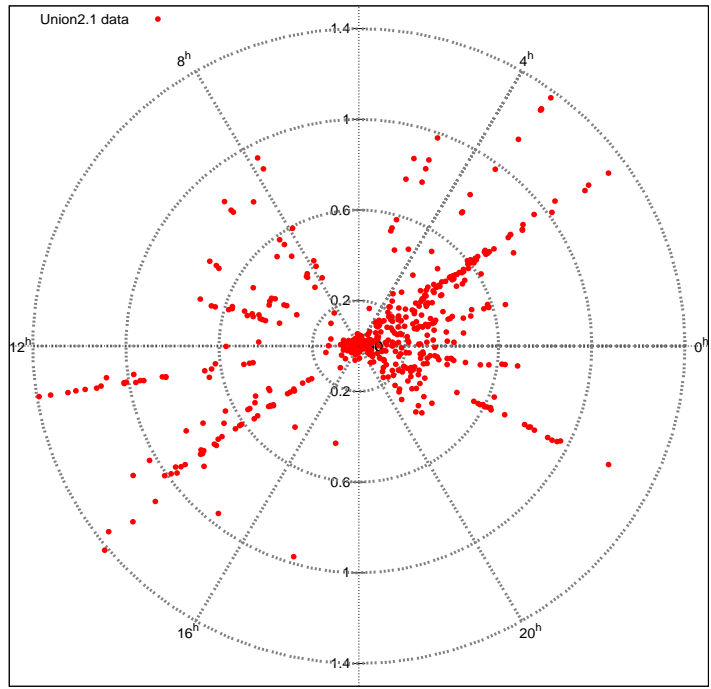}
 \caption{Pie diagram of the 580 SNIa of the {\em Union2.1} set.}
\end{figure}

\subsection{Quantifying the Hubble flow}
The basic procedure by which the Hubble expansion is traced
observationally is through the so-called distance modulus, $\mu$,
defined by:
\begin{equation}
\mu=m-M=5\log{d_L}+25
\label{eq4}
\end{equation}
where $m$ and $M$ are the apparent and absolute magnitude of the SNIa,
and $d_L$ its luminosity distance, given for a
flat geometry ($\Omega_k=0$) by:
\begin{equation}
d_L=c(1+z)\int_0^z \! \frac{\mathrm{d}x}{H(x)}\; .
\label{eq2}
\end{equation}
with $H(z)$ the so-called Hubble parameter, defined using the 1st
Friedmann equation, by:
\begin{equation}
H(z)=H_0\sqrt{\Omega_{m,0}(1+z)^3+\Omega_{k,0}(1+z)^2+
\Omega_\Lambda\exp\left(3\int_0^z \! \frac{1+w(x)}{1+x} \, \mathrm{d}x\right)}
\label{hubble}
\end{equation}
with $H_0$ the Hubble constant and 
$\Omega_{m,0},\ \Omega_{k,0},\ \Omega_{\Lambda}$ the 
fractional densities of matter, curvature and the cosmological
constant or DE respectively at the present epoch, while $w$ is the equation
of state parameter of Dark Energy. It is evident that the luminosity
distance depends strongly on the cosmological parameters. Different
DE models reflect to the functional form of the $w(x)$. The QDE model
(Quintessence DE), that we will use in the current study, assumes
a constant equation of state parameter, $w$, that can admit values
different than $-1$.

The assumption of the isotropic Hubble expansion boils down in having
a Hubble parameter $H(z)$ being a function only of $z$ and not of
direction. Therefore, different radial directions should provide
statistically equivalent $H(z)$. The approach we use in this analysis
is exactly to investigate different radial directions, or solid
angles, in order to confirm or not the independence of $H(z)$ on
direction. The criterion of equivalence in the Hubble expansion among
the different solid angles will be whether the resulting cosmological
parameters, ($\Omega_{m,0}, w$), provided by fitting the data to the above
Hubble expansion models, are statistically consistent among
them and among the solution using the whole ($z>0.02$) {\em Union2.1}
compilation.

\subsection{The $\chi^2$-minimization procedure}
In order to fit the cosmological parameters of eq.(\ref{hubble}) we
minimize the difference between the SNIa observed distance moduli and
the theoretical ones via eq.(\ref{eq4}). To this end we use a $\chi^2$
minimization procedure outlined below.

Suppose we have a measurement of the distance modulus
$\mu_{obs}(z_i)$ with uncertainty $\sigma_\mu$, which comes with its
redshift $z_i$. The theoretically
expected distance modulus for the same redshift $z_i$ is
$\mu_{th}(z_i,\textbf{p})$, with $\textbf{p}\equiv {w,\Omega_{m,0}}$ being
the model's free parameters. Now suppose that we have $N$
$\mu_{obs}(z_i)$ independent measurements. The total probability of
obtaining this entire set of these $N$
data points is equal to the product of the probability of each data
point, so
:
\begin{equation}
P_{\rm tot}=\prod\limits_{i=1}^N P(z_i)=\left[\prod\limits_{i=1}^N
  \frac{1}{\sigma _i \sqrt{2\pi}}\right]\exp{\left[-\frac{1}{2}
    \sum\limits_{i=1}^N\left(\frac{\mu_{obs}(z_i)-\mu_{th}(z_i,\mathbf{p})}
               {\sigma_i}\right)^2\right]}
\end{equation}
If we want to find the maximum probability we have to minimize the sum
in the exponential term of $P_{\rm tot}$, and therefore, this
quantity is defined as:
\begin{equation}
\chi^2=\sum\limits_{i=1}^N\left(\frac{\mu_{obs}(z_i)-
  \mu_{th}(z_i,\mathbf{p})}{\sigma_i}\right)^2
\label{eq 1}
\end{equation}
the minimum value of which, $\chi^2_{min}$, provides the maximum
probability of $P_{\rm tot}$. Thus, since the theoretical expected
$\mu_{th}(z_i,\textbf{p})$, depend on a set of free parameters, which
correspond to the elements of the vector \textbf{p} [in our case ${\bf
  p}=(\Omega_{m,0}, w)$], we can test for
which values of these parameters we get the maximum probability. In
our current analysis we ignore the covariance matrix of the errors in
the observed SNIa distance moduli and use the root-mean-square of the
diagonal elements, provided with the {\em Union2.1} catalogue. Also,
we use the approach of Nesseris \& Perivolaropoulos (2006) by
which we do not need to impose an {\em a priori} value of the Hubble
constant.

The best fit value of the free parameters, $\mathbf{p_0}$, are provided 
for the $\chi^2_{min}$. When the parameters differ from these
values, $\mathbf{p}\neq \mathbf{p_0}$, then the $\chi^2$ increases, so
$\Delta\chi^2=\chi^2-\chi^2_{min}>0$. Limits of $\Delta\chi^2$ that
depend on the number of the fitted parameters $N_f$, define confidence
regions that contain a certain fraction of the probability
distribution of \textbf{p}'s. For our case of $N_f=2$,
 the 1, 2 and 3 $\sigma$ confidence regions correspond to
 $\Delta\chi^2=2.3, 6.17$ and 11.83, respectively.

\section{Results}
Using the whole ($z>0.02$) {\it Union2.1} sample we derive, as expected,
the already published QDE model constrains of:
\begin{equation}\label{eq:tot}
\Omega_{m,0}=0.282\pm 0.03 \;\; {\rm and} \;\; w=-1.013\pm 0.083 \;\;
{\rm with} \;\; \chi^2/dof=0.9567
\end{equation}
fully consistent with those of Suzuki et al. (2012).
Now, we proceed with the main scope of our work which is to identify
possible anisotropies of the Hubble flow. 

\subsection{Dividing the sky into 9 solid angles}
As a first test we divide the celestial sphere in nine fully
independent solid angles, shown in Figure 2, so that each has
a similar number of SNIa.
We then fit the QDE cosmological parameters in each, by using the
previously described minimization procedure. 
Doing so, we identified one sky region with a distinctly different
$\Omega_{m,0}-w$ solution with respect to all other regions of the sky as
well as with respect to the whole {\em Union2.1} sample together. 
This region, which we call Group X, contains 82 SNIa and has
galactic coordinates within: $35^\circ<l<83^\circ$ and
$-79^\circ<b<-37^\circ$ (delineated within thick lines in Figure 2). 
We have verified that the SNIa redshift distribution in this region
does not present any ``peculiarity'' with respect to the overall. In
fact, although the redshift distributions of the separate regions show 
different levels of consistency with the overall {\em Union2.1}
redshift distribution, the Kolmogorov-Smirnov significance of
which vary from ${\cal  P}\sim 10^{-7}$ up to $\sim 0.13$, 
that of the Group X is the most statistically consistent with the
overall (${\cal P}\simeq 0.13$). Thus a ``peculiar'' redshift
distribution is not the cause of the Group X's $\Omega_{m,0}-w$
solution behaviour.

\begin{figure}[t!]
\centering
\includegraphics[width=1.05\textwidth,height=7.5cm]{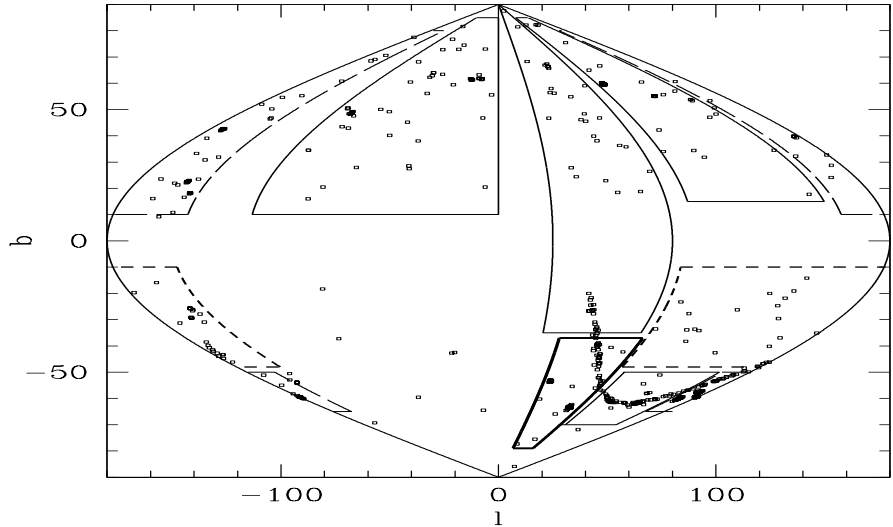}
\caption{
The celestial sphere in equal area projection with 8 separate regions
delineated (the 9th region is the rest of the sky). 
The area delineated with thick lines corresponds to Group-X. Dashed
lines are used when a region folds over into the Galactic
longitude direction.}
\end{figure}

\begin{figure}[!t]
\center
  \includegraphics[width=0.48\textwidth,height=5.8cm]{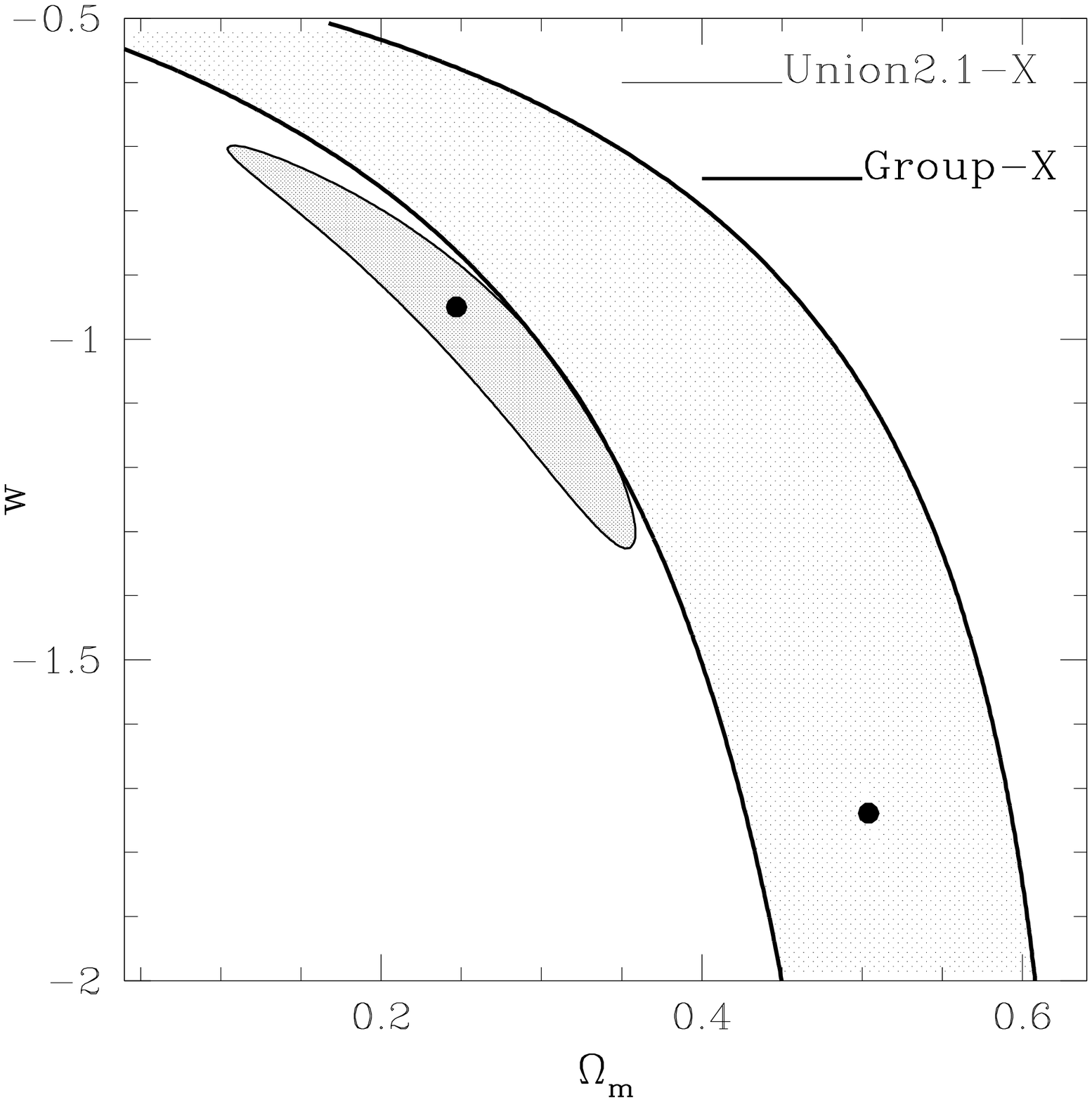}
\hfill
  \includegraphics[width=0.48\textwidth,height=5.8cm]{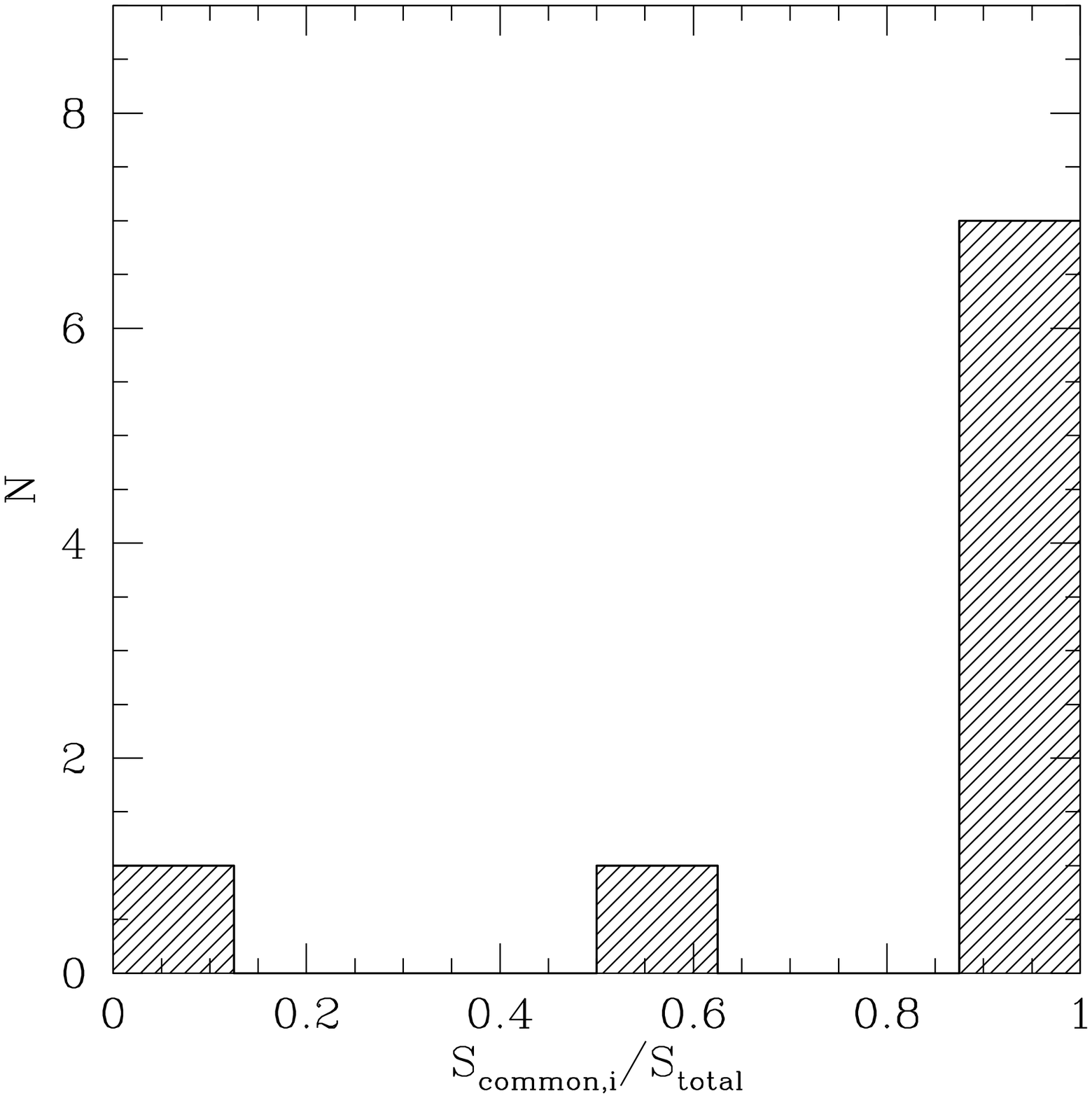} 
\vspace*{8pt}        
\caption{{\it Left panel:} 1$\sigma$
  contour plots of Group X (thick contour and light greyscale, 82
  SNIa) and for the rest
  of the {\em Union2.1} data (thin contour and dark greyscale, 464
  SNIa). {\it Right panel:}
Distribution of the common 1$\sigma$ contour area (between
  each of the 9 individual subsamples and the whole SNIa sample)
  divided by the 1$\sigma$ area of the entire sample.
\label{fig17}}
\end{figure}

In the left panel of Figure \ref{fig17} we compare the 1$\sigma$
contour region of Group X with that of the of the rest of the
{\em Union2.1} sample and as it can be seen their respective
1$\sigma$ contour regions have no common area, indicating a relatively
significant difference. Quantitatively, the difference of the latter
best-fit solution with respect to that of Group X
is of $\Delta\chi^2\simeq 4.3$ (ie., the probability of being different is
$\sim 90\%$).

In order to visualise the uniqueness of the Group X
behaviour we plot in the right panel of Figure \ref{fig17} 
the fraction of the QDE 1$\sigma$ contour area, based on the whole 
{\em Union2.1} (excluding the Group X), which is common with
that based on every one of the 9 independent subsamples of
SNIa.
We observe that the 1$\sigma$ solution space of the entire sample
is fully (100\%) encompassed within almost each of the other 8 subsamples
(with one exception which is at 60\%; the corresponding region is delineated
with short-dashed lines in Figure 2), besides Group X which is at 0\%.

It is interesting to note that Group X is near the CMB dipole
anti-apex direction. Furthermore, many studies have consistently been
finding the largest deviations from an isotropic expansion occurring
near this direction. For example, as early as 2001 and using only 79
SNIa, Kolatt \& Lahav (2001) found a Hubble diagram dipole pointing
towards $(l, b)\simeq (80^{\circ}, -20^{\circ})$.
Antoniou \& Perivolaropoulos (2010) using the {\it Union2} SNIa sample
and separately Cooke \& Lynden-Bell (2010), using the slightly earlier
{\em Union} sample, also found 
a preferred axis of minimum acceleration in the same general direction
but at a low significance level.
Yang, Wang \& Chu (2013) using the {\it Union2.1} SNIa data
 also found a low significance preferred direction of the
accelerating expansion. Finally, Javanmardi et
al. (2015), found as the most discrepant direction of the
Hubble expansion that centered on $(l, b)\simeq (67.5^{\circ},
-66.4^{\circ})$, which indeed coincides with our Group X.

\subsection{Possible systematic effects}
The statistically important ``erratic'' behaviour of Group X with
respect to the rest
of the SNIa could be due to a variety of reasons, among which an
intrinsic anisotropy of the Hubble expansion, 
a large bulk flow in this part of the sky or
some unknown systematic observational error (see for example Heneka et
al. 2014).
In this respect we remind the reader that the SNIa distance moduli,
provided in the {\em Union2.1} compilation,
have been derived assuming a global correction for individual deviations from
the average SNIa light-curve and from the mean color. 
As highlighted in Javanmardi et al. (2015), the
most robust approach for testing isotropy would be to find the values for
the ``stretch'', color and host-galaxy parameters 
for each patch of the sky separately. However, we adopt the simplified
approach of the global correction, as in Javanmardi et al. (2015).

Below we present our analysis of two possible systematic observational
effects that could be the cause of the ``erratic''  behaviour of Group X.

\subsubsection{Galactic absorption}
\begin{figure}[!t]
\center
  \includegraphics[width=0.85\textwidth,height=8cm]{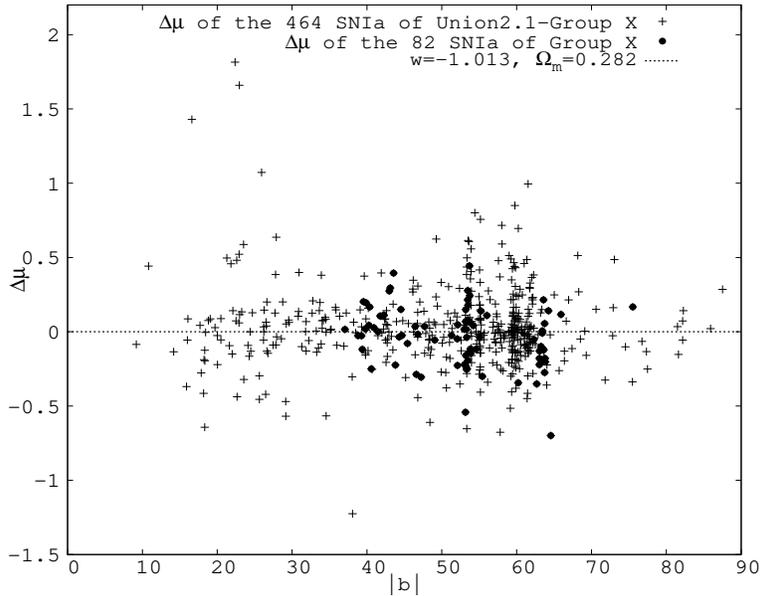}
\vspace*{8pt}        
\caption{
Distance moduli deviation $\Delta\mu$, with respect
to that of $(\Omega_{m,0}, w)=(0.282, -1.013)$, as a function of galactic
latitude $|b|$ for the 546 {\it Union2.1} SNIa (crosses) and for
the 82 SNIa of Group X (filled dots).}
\label{dm-b}
\end{figure}

We test whether the erratic behaviour of Group X could be due to
inadequately treating the Galactic absorption. To this end and for
each SNIa we plot in Figure \ref{dm-b} the distance modulus deviation from
that expected in the concordance model, ie.,
$\Delta\mu=\mu_{obs}(z_i)-\mu_{th}(z_i,\mathbf{p})$ with
$\mathbf{p}\equiv (w,\Omega_{m,0})\equiv (-1.013, 0.282)$, as a function
of the Galactic latitude $|b|$.

As it is illustrated in Figure \ref{dm-b}, the SNIa distance moduli of
Group X (in blue) do not show a distinct behaviour as a function of
$|b|$. However, we see that the SNIa with the largest
$\Delta\mu$'s, i.e., the apparently faintest ones, are all at small
$|b|$'s, which strongly hints towards some at least SNIa having
magnitudes not adequately corrected, or overcorrected, for Galactic absorption. These
SNIa ({\em 1997k} with $b=16.6^{\circ}$, {\em 1997l} with
$b=22.4^{\circ}$ and {\em 1997o} with $b=22.9^{\circ}$)
are also among the nearest to the plane of the Galaxy.

\begin{figure}[!t]
\center
\includegraphics[width=0.9\textwidth,height=9cm]{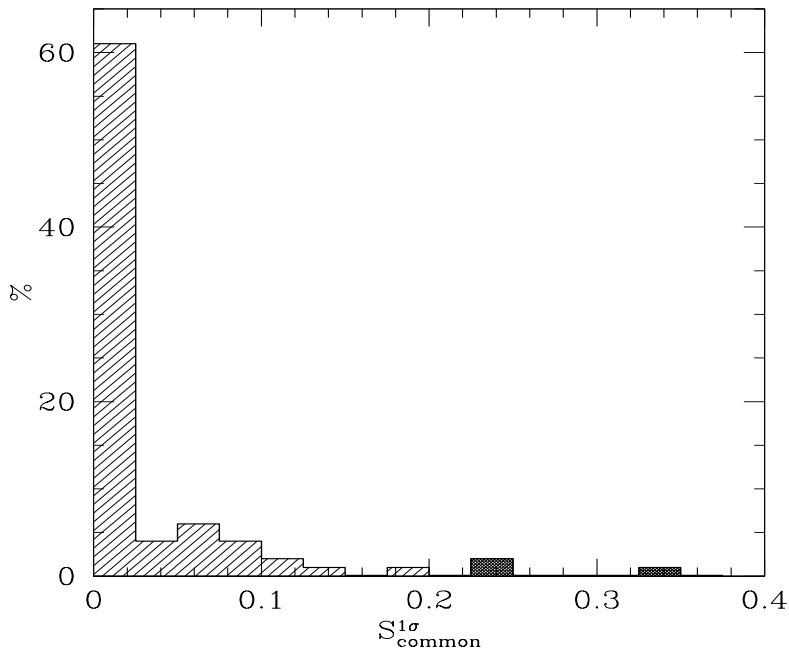}
\vspace*{8pt}        
\caption{Fraction of the {\em Union2.1} (excluding Group X) 
QDE 1$\sigma$ solution area which is common with the corresponding
area of Group X excluding one by one each Group X SNIa from the fitting
procedure. Marked by dark shade are the three SNIa with the largest
effect, the exclusion of which recovers the consistency of the Hubble
expansion of Group X
with that of the overall {\em Union2.1} sample.}
\end{figure}

\subsubsection{SNIa Outliers ?}
We have investigated the possibility whether an erratic behaviour of one or a few SNIa 
could cause the mentioned effect. To this end we systematically
exclude, one by one, each of the SNIa of Group X and check whether the
QDE 1$\sigma$ contour area of the remaining Group X increases its consistency with the 
{\em Union2.1} solution (excluding the Group X).
We find that although the large majority of the single SNIa do not
have any major contribution to the erratic behaviour of Group X, 
we do find that some individual SNIa have a very large effect on the
Group X solution, as can be seen in Figure 5 where we plot the
fraction of the overall {\em Union2.1} (excluding the Group X) 1$\sigma$
contour area covered by the corresponding Group X 1$\sigma$ 
contour area when excluding
one SNIa at a time. The three cases that show a $>20\%$ effect are
shown in dark shade. 
The largest effect, that of a $\sim 35\%$ recovery of the common
1$\sigma$ solution space, comes from excluding the {\em 03D4cx} SNIa at
$z=0.949$. The next most important effect
comes from {\em g050} at $z=0.613$, which by excluding it 
results in a $\sim 25\%$ recovery of the common
1$\sigma$ solution space, while the third most important effect, again
at a $\sim 23\%$ level, comes from the {\em 2005hv} at $z=0.1776$.
Most importantly, the joint effect of excluding all three 
``erratic'' SNIa is the increase of the consistency from 0\% to more than
93\%, thus annulling the tension
between the Group X and the rest of the {\em Union2.1} sample solution.

We conclude that the identified inconsistency of the Hubble expansion
of the Group-X SNIa appears to be dominated by the erratic behaviour
of only three
out of the 82 SNIa, indicating the sensitivity of the Hubble
expansion solution to outliers when small numbers of SNIa are
investigated.

Excluding from the overall analysis the above three ``erratic'' SNIa as well
as the 3 SNIa at low galactic
latitudes, which are suspect of an inadequate Galactic absorption
correction, we find:
\begin{equation}
\Omega_{m,0}=0.262^{+0.0313}_{-0.0296} \;\; {\rm and} \;\; w=-0.973\pm 0.077 \;\;
{\rm with} \;\; \chi^2/dof=0.909
\end{equation}
Comparing with the solution using all the {\em Union2.1} set
(eq. \ref{eq:tot}) we see a small but notable difference of the best fit
$\Omega_{m,0}, w$ parameter values, while the $\chi^2/dof$ is also lower.
 
\begin{figure}[t!]
\centering
  \includegraphics[width=0.99\textwidth,height=9.8cm]{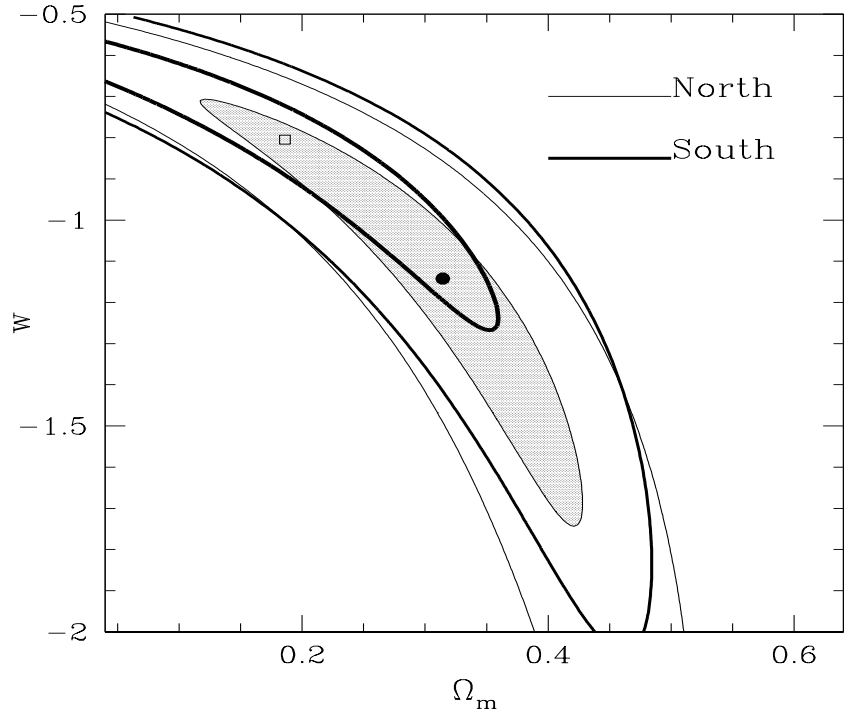}
\vspace*{8pt}        
\caption{
The separate Galactic hemisphere solution: Thick lines correspond to the
Southern G.H. while the greyscale to the Northern G.H. with their respective
best fit values shown as empty and filled dots, respectively.}
\label{fig4}
\end{figure}

\subsection{Separate Galactic hemispheres solutions}
We now consider, as a final test of possible Hubble expansion
anisotropies, the southern and northern
Galactic hemispheres separately, excluding the 6 ``erratic'' SNIa
identified previously. We therefore use 332 and 208 SNIa,
respectively for the southern and northern hemisphere. 
The results are:

\begin{itemize}
\item South:
$w=-0.805\pm 0.084$, $\Omega_{m,0}=0.186^{+0.0508}_{-0.0468}$, $\chi^2_{min}/dof=0.9492$
\item North:
$w=-1.144^{+0.1325}_{-0.1440)}$, $\Omega_{m,0}=0.315^{+0.0405}_{-0.0377}$, $\chi^2_{min}/dof=0.8500$
\end{itemize}
It can be seen that the best-fit QDE model parameters for the two hemispheres are
quite different, with a $\Delta\chi^2=1.27$,
but within the 1$\sigma$ uncertainty contour range, as seen in Figure 6.
Therefore this difference indicates some tension between the two
hemisphere solution but does not appear to constitute evidence of
a significant anisotropy.

\section{Conclusions}
We have searched for possible anisotropies of the Hubble
expansion, using the {\em Union2.1} SNIa data. Our approach was to
investigate separately the Hubble expansion traced by SNIa 
in 9 solid angles, covering the whole sky, as well as separately in the two
Galactic hemispheres.
We have identified only one particular sky region, with galactic coordinates
$35^\circ<l<83^\circ$ \& $-79^\circ<b<-37^\circ$ (near
the CMB dipole anti-apex direction),
which provides a relatively significant different $\Omega_{m,0}-w$
solution with respect to the rest of the {\em Union2.1} compilation 
or any other subsample of similar size that we have analysed. 
Our investigation for the possible causes of this result has shown
that probably it should not be attributed to an intrinsic 
anisotropy in the Hubble flow but rather on the erratic behaviour of
only three SNIa, indicating the
sensitivity of the measured Hubble expansion to outliers when
relatively small numbers of SNIa are investigated.
We have also found that the 3 most deviant SNIa distance moduli are
located at very low-galactic latitudes, suggesting an inadequate
Galactic absorption correction. Excluding the six problematic SNIa
from the {\em Union2.1} sample results in a notable
change the QDE model free parameters but not at a significant level.
Finally, we have also found a quite different QDE parameters fit
between the North and South Galactic hemispheres, but not 
significant enough to assert the detection of a Hubble expansion
anisotropy.


\begin{thebibliography}
\bibitem[Amanullah(2010)]{}Amanullah, R., Lidman, C., Rubin, D.,
  et al. 2010, \apj,  \textbf{716}, 712
\bibitem[Antoniou(2010)]{} Antoniou, A., Perivolaropoulos, L. 2010, JCAP, \textbf{12}, 012

\bibitem[Appleby(2015)]{} Appleby, S., Shafieloo, A. and Johnson, A. 2015, ApJ, \textbf{801}, 76

\bibitem[Barbon(1973)]{} Barbon, R., Ciatti, F., \& Rosino, L. 1973, A\&A, \textbf{25}, 241

\bibitem[Betoule(2014)]{}Betoule, M., Kessler, R., Guy, J., et
  al. 2014, A\&A, {\bf 568},  22

\bibitem[blomqvist(2008)]{}Blomqvist, M., Mortsell, E., \& Nobili,
  S. 2008, JCAP, \textbf{6}, 27

\bibitem[Bonvin(2006)]{}Bonvin, C., Durrer, R., \& Kunz, M. 2006, PRL, \textbf{96}, 191302

\bibitem[Cooke(2010)]{} Cooke, R., Lynden-Bell, D. 2010, MNRAS,
  \textbf{401}, 1409

\bibitem[Ghirlanda(2006)]{} Ghirlanda, G., Ghisellini, G. and Firmani, C. 2006,
  NJPh, {\bf 8}, 123

\bibitem[Gupta(2010)]{}Gupta, S., \& Saini, T. D. 2010, MNRAS, \textbf{407}, 651

\bibitem[Heneka(2014)]{}Heneka, C., Marra, V., \& Amendola, L. 2014,
  MNRAS, \textbf{439},  1855

\bibitem[Javanmardi(2015)]{} Javanmardi, B., Porciani, C., Kroupa, P.,
  Pflamm-Altenburg, J. 2015, ApJ, \textbf{810}, 47

\bibitem[Kolatt(2001)]{} Kolatt, T. S., Lahav, O. 2001, MNRAS, \textbf{323},
  859

\bibitem[Kalus(2013)]{} Kalus, B., Schwarz, D. J., Seikel, M., \&
  Wiegand, A. 2013, A\&A, \textbf{553}, A56

\bibitem[Kowal(1968)]{} Kowal, C. T. 1968, ApJ, \textbf{73}, 1021

\bibitem[Kowalski(2008)]{}Kowalski, M., Rubin, D., Aldering, G., et
  al. 2008, ApJ,  \textbf{686}, 749

\bibitem[Ma(1996)]{} Ma, Y., \& Pan, J. 2014, MNRAS, \textbf{437}, 1996

\bibitem[Mariano(2012)]{} Mariano, A., Perivolaropoulos, L. 2012, PRD, \textbf{86}, 083517

\bibitem[Melnick(2000)]{} Melnick, J., Terlevich, R., Terlevich, E. 2000, MNRAS, \textbf{311}, 629

\bibitem[Nesseris(2006)]{} Nesseris, S., Perivolaropoulos, L. 2006, PRD, {\bf 73}, 103511 

\bibitem[Perlmutter(1999)]{} Perlmutter, S., et al. 1999, ApJ, \textbf{517}, 565

\bibitem[Ade(2014)]{} Planck Collaboration (P. Ade \textit{et. al}.)
  2014, A\&A,  \textbf{571}, A1

\bibitem[Plionis(2011)]{} Plionis, M., Terlevich, R., Basilakos, S., Bresolin,
  F., Terlevich, E., Melnick, J., Chavez, R. 2011, MNRAS, \textbf{416}, 2981

\bibitem[Riess(1998)]{} Riess, A. G., et al. 1998, AJ, \textbf{116}, 1009

\bibitem[Schwarz(2007)]{}Schwarz, D. J., \& Weinhorst, B. 2007, A\&A,
  \textbf{474}, 717

\bibitem[Suzuki(2012)]{} Suzuki N. et al. 2012, ApJ, \textbf{746}, 85

\bibitem[Wu(1999)]{} Wu, K. S.,  Lahav, O., \&  Rees, M. J. 1999,
  Nature, \textbf{397}, 225


\bibitem[Yang(2014)]{} Yang X. F., Wang F. Y., Chu Z. 2014, MNRAS,
  \textbf{437}, 1840






\end{thebibliography}
\end{document}